\documentclass[a4paper,fleqn]{cas-sc}
\usepackage{booktabs}
\usepackage{multirow}
\usepackage[numbers]{natbib}
\usepackage{etoolbox}

\def\tsc#1{\csdef{#1}{\textsc{\lowercase{#1}}\xspace}}
\tsc{WGM}
\tsc{QE}
\tsc{EP}
\tsc{PMS}
\tsc{BEC}
\tsc{DE}

\usepackage{hyperref}
\usepackage{tabularx}
\usepackage[utf8]{inputenc}
\usepackage{fancyhdr}
\usepackage{multirow}
\usepackage{float}
\usepackage{amsmath,amsfonts,amssymb,amsthm}
\usepackage{enumerate} 
\everymath{\displaystyle}
\usepackage{caption}
\usepackage{tikz}
\usepackage{graphicx} 
\usepackage{bm}
\usepackage{sidecap}
\usepackage{wrapfig} 
\usepackage{listings}
\theoremstyle{definition}
\usepackage{hyperref}
\usepackage{xcolor}
\usepackage[normalem]{ulem}
\usepackage{amssymb}
\usepackage{algorithm}
\usepackage[noend]{algpseudocode}
\usepackage{amsmath}
\usepackage[toc,page]{appendix}
\usepackage{subcaption}



\usepackage{booktabs}
\usepackage{multirow}
\usepackage{adjustbox}
\begin{document}
\let\WriteBookmarks\relax
\def\floatpagepagefraction{1}
\def\textpagefraction{.001}
\shorttitle{Real-time and adaptive anomaly detection algorithm for cyclostationary models}
\shortauthors{J. Witulska et~al.}

\title[mode=title]{Real-time and adaptive anomaly detection algorithm for cyclostationary models}

\author[1]{Justyna Witulska}[orcid=0000-0003-4351-0783]
\cormark[1]
\ead{justyna.witulska@pwr.edu.pl}
\credit{Conceptualization, Methodology, Software, Writing -- Original Draft}

\affiliation[1]{organization={Faculty of Pure and Applied Mathematics and Hugo Steinhaus Center,
                Wrocław University of Science and Technology},
                city={Wrocław},
                postcode={50-370},
                country={Poland}}

\author[2]{Tomasz Barszcz}
\ead{tbarszcz@agh.edu.pl}

\affiliation[2]{organization={AGH University},
                city={Kraków},
                postcode={30-059},
                country={Poland}}

\author[3]{Ireneusz Jabłoński}
\ead{ireneusz.jablonski@b-tu.de}

\affiliation[3]{organization={Faculty of MINT, Brandenburg University of Technology},
                city={Cottbus},
                postcode={03046},
                country={Germany}}

\author[1]{Agnieszka Wyłomańska}
\ead{agnieszka.wylomanska@pwr.edu.pl}

\cortext[cor1]{Corresponding author}

\begin{abstract}
 This article introduces PeriodicCALM, an effective real-time anomaly detection framework designed for cyclostationary data streams. While classical cyclostationary processes feature periodically time-varying statistical properties, real-world signals often contain recurring impulsive components that conceal abnormal behavior. Existing real-time methods for  struggle with these dynamics, frequently misinterpreting phase-dependent variability as non-cyclic anomalies and causing excessive false alarms. To address this, PeriodicCALM incorporates cycle-dependent variability to systematically ignore regular cyclic impulses while accurately isolating genuine anomalies. Operating in real time with continuous retraining capabilities, the method adapts dynamically to evolving signal characteristics. Comparative evaluations against the baseline CALM framework using simulated data demonstrate significant improvements in detection accuracy and training efficiency, alongside a reduction in prediction latency. Furthermore, the practical utility of PeriodicCALM is validated on real-world vibration signals collected from a compressor monitoring system.
\end{abstract}



\maketitle

\section{Introduction}
\label{sec:intro}
The primary goal of this article is to propose a  methodology, termed PeriodicCALM, for continuous, real-time anomaly detection in data with cyclostationary behavior. Second-order cyclostationary processes (for simplicity called later cyclostationary) are stochastic models whose statistical characteristics, such as mean or autocovariance function (ACVF), vary periodically over time \cite{napolitano1,hurd2007periodically}. They arise naturally in many real applications, including condition monitoring \cite{antoni2016cyclostationarity,ZULAWINSKI2024111367}, telecommunication \cite{7520421}, and physiological monitoring \cite{KAZEMI201479}. 
A common representative of the cyclostationary models is the periodic autoregressive (PAR) model, which captures cyclic temporal dependence while allowing parameters to change periodically over time \cite{hurd2007periodically}. The second classical model exhibiting cyclostationary behavior consists of a deterministic periodic function combined with an additive stationary time series. In this article (for specific periodic function) we refer to this as the vibration signal model, as similar behavior, characterized by a specific deterministic periodic component, is observed in compressor vibration signals \cite{grzesiek2026impulsivity}. The analysis of classical cyclostationary models relies on the utilization of the ACVF and its examination in both the time and frequency domains \cite{zulawinski2025fractional}. A standard algorithm for identifying cyclostationary behavior in data is cyclic spectral coherence (CSC), which constitutes a double Fourier transform of the ACVF, with its estimator founded upon the classical sample ACVF \cite{ANTONI2007597, cui2024spectral}.

The literature also considers cyclostationary models with additional disturbances. Classic examples include PAR models with additive noise or  with additive outliers \cite{10714997,zulawinski04052025,ZULAWINSKI2023115131}. Behaviors corresponding to such disturbed models are also observed in real-world applications, for instance in data describing air quality parameters \cite{emp_study2}, flows of rivers \cite{SARNAGLIA20102168}, as well as in vibration signals from mechanical systems \cite{ZULAWINSKI2024111367}. Algorithms designed for disturbed cyclostationary models (particularly in scenarios involving impulse noise or outliers) frequently rely on robust estimators of the ACVF. Within this domain, robust algorithms for computing CSC map \cite{ZULAWINSKI2024111367} warrant particular attention, as they have found widespread application in the field of condition monitoring. 

In general, two main approaches can be adopted when dealing with models affected by additional disturbances. The first scenario involves adapting existing methodologies  to explicitly account for the presence of these disturbances, like mentioned above robust versions CSC maps. However, a major drawback of such robust methods is their often substantially increased computational complexity, which can result in long computation times and thus limit their applicability in real-time applications (especially in online monitoring). The second scenario focuses on identifying and potentially removing anomalies prior to applying standard classical models for further analysis. In this paper, we adopt the latter approach.

Anomaly detection techniques play a critical role in monitoring and maintaining complex systems, especially within environments generating large volumes of continuous, heterogeneous data. These techniques are applied across diverse sectors, including cybersecurity, healthcare, industrial operations, and smart environments \cite{fahrmann2024anomaly, klots2025intelligent, yang2025research, gaggero2025artificial, ali2025innovative, ali2025innovative}. 
A widely used classification divides anomaly detection methods into three categories: point anomalies (individual observations that significantly deviate from expected behavior), contextual anomalies (data points whose abnormality depends on the surrounding context), and collective anomalies (sequences of related observations that are anomalous only when considered as a whole) \cite{klots2025intelligent, latif2025gat, khettaf2026deep}.

A broad spectrum of methods has been proposed for anomaly detection in both univariate and multivariate data~\cite{Raj2024ACS}. While these approaches have demonstrated considerable effectiveness across a range of applications, their practical applicability to multivariate data is often constrained by computational and operational requirements, with many methods achieving optimal performance in offline rather than real-time settings \cite{raval2025anomaly}. From a methodological perspective, anomaly detection techniques can be broadly classified into four principal categories: statistical methods, probabilistic approaches, machine learning-based techniques, and hybrid methods that combine complementary detection paradigms~\cite{samariya2023comprehensive, pyod_package, ghost, chen2025pyod, vijay2020timeseries, xu2021anomaly, alla2019beginning, mao2021toward, abououf2023explainable}.  

Since many of these methods rely on extensive feature engineering, labeled data, or computationally intensive training, their practical deployment in real-time, large-scale environments is constrained~\cite{habeeb2019real}. To address these challenges, we recently proposed an unsupervised method, called CALM, a nonparametric algorithm based on kernel density estimation and bootstrap-based thresholding, designed for anomaly detection at the individual level in real-time applications~\cite{witulska2026realtime}. However, CALM assumes independence between observations, which limits its applicability to cyclostationary signals where natural oscillations may be misclassified as anomalies, leading to significantly increased false positive rates. This highlighted the need for further development of methodologies tailored to this type of data.

Hence, this research addresses the gap regarding computationally efficient, unsupervised anomaly detection methods capable of accounting for the inherent temporal dependencies and periodic structure of cyclostationary signals, while maintaining reliable detection performance in real-time applications. To address this gap, PeriodicCALM is proposed, an extension of the CALM framework that incorporates cycle-dependent variability modeling through adaptive envelope construction. The main contributions of this study are: (i) the proposition of a cyclic envelope construction method that models natural variability within signal cycles using local quantiles and variance adjustment, enabling the distinction between normal cyclic behavior and genuine anomalies; (ii) the development of an envelope-based post-processing filtering mechanism that validates detected anomalies against learned cycle patterns, significantly reducing false alarms in periodic data while preserving the simplicity and computational efficiency of the original CALM approach; (iii) the evaluation of the proposed methodology on cyclostationary data (simulated from two models presented in Section~\ref{sec:models}) with non-cyclic impulse contamination, demonstrating improved detection performance compared to the baseline approach. The effectiveness of PeriodicCALM was evaluated using well-established performance metrics commonly applied to classification algorithms (see Section~\ref{sec:efficacy} for details), where anomaly detection is formulated as a binary classification problem with anomalies representing the minority class; (iv) the demonstration of the practical application to real-world vibration data from a compressor monitoring system, confirming the utility of the method for condition monitoring in industrial settings.

The article is structured as follows. Section~\ref{sec:models} introduces the theoretical background of models used. Section~\ref{sec:methodology} provides a description of the proposed methodology, including the baseline CALM algorithm and the enhanced PeriodicCALM extension. Section~\ref{sec:simulation} presents the performance assessment for simulated data from both models. Section~\ref{sec:real_data} demonstrates the application to real compressor data. Finally, Section~\ref{sec:conclusions} summarizes the research and outlines future work directions.

\section{Cyclostationary model with additive outliers}
\label{sec:models}

In this section, we consider two cyclostationary models with additive outliers represented by additive impulses. The general form of the considered model is as follows:
\begin{equation}
Y_t = X_t + \tilde{B}_t, \qquad t \in \mathbb{Z}.
\label{eq:general_signed}
\end{equation}
In the above $\tilde{B}_t=\operatorname{sign}(X_t) \cdot B_t$ or $\tilde{B}_t= B_t$, depending on whether the added impulses take positive and negative values or only positive values, respectively, where 
$\operatorname{sign}(\cdot)$ denotes the sign function, defined as:
\begin{equation}
\operatorname{sign}(x) =
\begin{cases}
-1, & x < 0,\\
0, & x = 0,\\
1, & x > 0.
\end{cases}
\label{eq:sign_function}
\end{equation}
In Eq. (\ref{eq:general_signed}), the time series $\{B_t\}$ represents the sequence of independent and identically distributed (i.i.d.) time series such that for each $t$, $B_t \sim B$, where: 
\begin{equation}
B = K \cdot W,
\label{eq:impulse_model}
\end{equation}
where, $K$ is a uniformly distributed random variable on the interval $(a,b)$, where $b > a > 0$, and $W$ is a Bernoulli distributed random variable such that $P(W = 1) = p$, $P(W = 0) = 1 - p$, $p \in [0, 1]$. We assume that $K$ and $W$ are independent. Let us note that parameters $a$ and $b$ control the amplitude of the additive disturbances, while $p$ dictates their frequency of occurrence.

In both models, $\{X_t\}$ is a cyclostationary time series, i.e. a time series for which the mean and autocovariance functions are periodic in time. Precisely, the time series $\{X_t\}$  is cyclostationary with period $T > 0$ if for all $t, h \in \mathbb{Z}$ it satisfies the following:
\begin{equation}
\mathbb{E}(X_t) = \mathbb{E}(X_{t+T}), \quad cov(X_t, X_{t+h}) = cov(X_{t+T}, X_{t+h+T}),
\label{eq:cyclostationary}
\end{equation}
where $T$ is the smallest integer for which these conditions hold~\cite{hurd2007periodically}. 

In the following subsections, we discuss two  models of $\{X_t\}$ that fulfill the condition (\ref{eq:cyclostationary}). The first one is the periodic autoregressive (PAR) model, for which the mean function is constant (zero), and the autocovariance function is periodic in time, while the second one, which we call the compressor vibration model, has a periodic mean function, and its autocovariance function is independent of time point $t$. In the literature, the first model is called the cyclostationary model of order two, while the second one is referred to as cyclostationary model of order one. In our analysis, we also make different assumptions regarding the sign of the impulsive component $\tilde{B}_t$. In the first model (PAR), the anomalies may be either positive or negative, with their sign determined by the sign of the underlying process. In this case $\tilde{B}_t=\operatorname{sign}(X_t) \cdot B_t$ in the model defined in (\ref{eq:general_signed}). While for the second model, the anomalies are restricted to be positive additive disturbances. Thus, in this case, we assume $\tilde{B}_t=B_t$. 

\subsection{Periodic autoregressive model}
\label{sec:model}

The classical example of the cyclostationary model that fulfills conditions (\ref{eq:cyclostationary}) is the PAR model defined as follows \cite{franses2002forecasting}:
\begin{equation}
X_t - s_1(t)X_{t-1} - \cdots - s_{\kappa}(t)X_{t-\kappa} = Z_t,
\label{eq:PAR}
\end{equation}
where the functions $s_1(\cdot), \ldots, s_{\kappa}(\cdot)$ are periodic with a period $T$, $\kappa$ is the order of the model, and $\{Z_t\}$ is a innovation series with standard deviation $\sigma$. In the classical version of the model, $\{Z_t\}$ is assumed to be Gaussian white noise.
 In our analysis, for simplicity, we assume $\kappa=1$. In such a case, we denote $s_1(t)=s(t)$. We recall that the unique bounded (in the sense of $l_2$ norm) solution of Eq. (\ref{eq:PAR}) is given by \cite{aaw29}:
\begin{eqnarray}\label{par1_rozw}
X_t=\sum_{j=0}^{\infty}S_{t-j+1}^{t}\xi_{t-j},
\end{eqnarray}
where $S_{k}^n=\prod_{r=k}^ns(r)$ (with the convention $S_k^n=1$ when $k>n$) if and only if $|P|=|s(1)s(2)...s(T)|<1$.

The cyclostationarity of the PAR(1) model translates into the cyclostationarity of the model defined in Eq. (\ref{eq:general_signed}). Indeed, it is easy to show that the time series $\{Y_t\}$ has a mean equal to zero and an autocovariance function given by:
\begin{eqnarray}\label{nowe1}
cov(Y_{t},Y_{t+h})=cov(X_t,X_{t+h})+\mathbb{I}_{h=0}\sigma^2_{B}= cov(Y_{t+T},Y_{t+h+T}),
\end{eqnarray}
where $\mathbb{I}_A$ is the indicator of the set $A$ and $\sigma^2_{B}=Var(B_t)$. 

\begin{figure}
    \centering
    \includegraphics[width=0.6\linewidth]{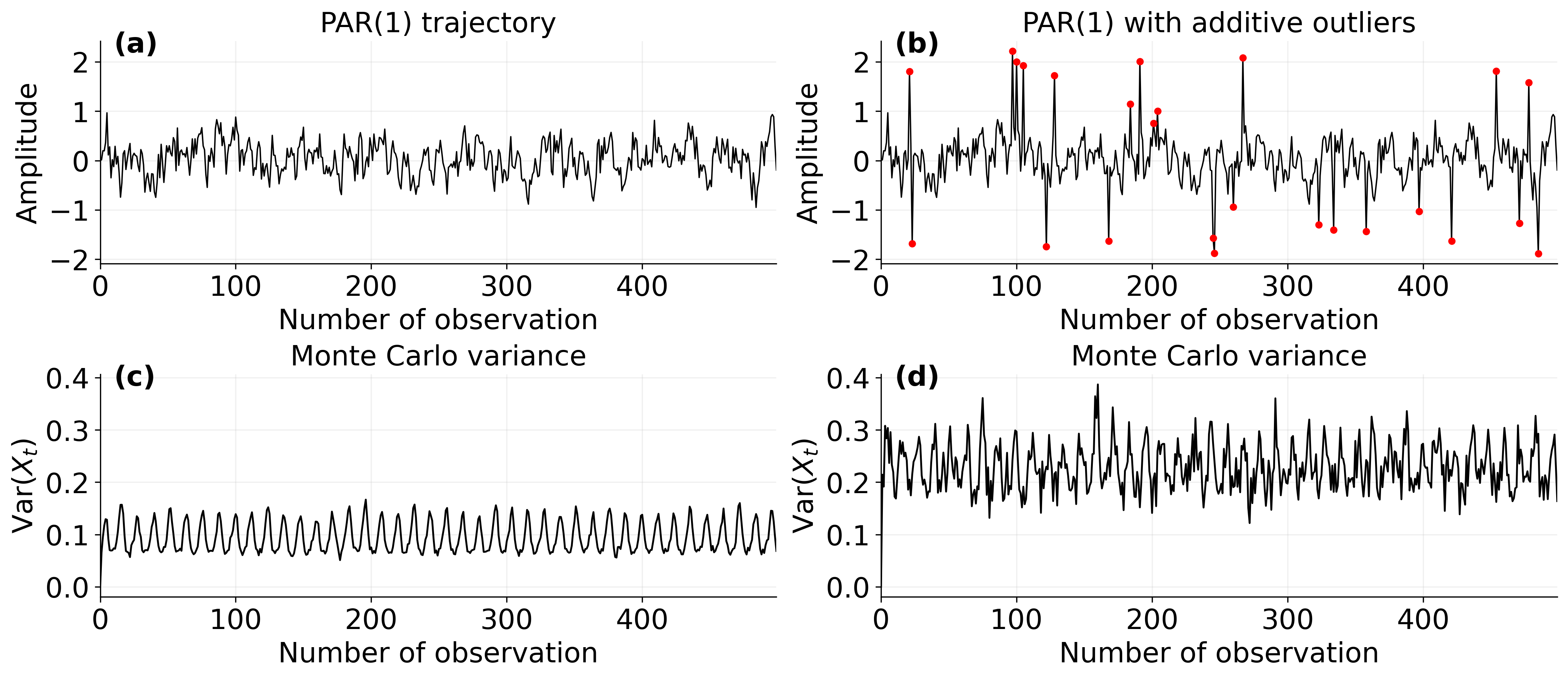}
    \caption{Example trajectories of (a): the PAR(1) model  and (b): its version with additive outliers described as $\tilde{B}_t=\operatorname{sign}(X_t) \cdot B_t$, together with the corresponding variances estimated from M=500 Monte Carlo simulations (respectively: (c) and (d)).}
    \label{fig:traj_par1}
\end{figure}

In Fig. \ref{fig:traj_par1} we demonstrate the example trajectory of the PAR(1) model with a periodic coefficient $s_1(t)=s(t) = 0.5 + 0.3 \cdot \sin(2\pi t / 12)$, period $T = 12$, noise standard deviation $\sigma = 0.25$, and sample size $N = 5000$ (left panel) and the corresponding PAR model with additive outliers $\tilde{B_t}=sign(X_t)\cdot B_t$ with $a=0.75$, $b=2$, and $p=0.05$ (right panel). On the bottom panels, we present the corresponding variance (i.e., autocovariance function for $h=0$) of both models calculated using the Monte Carlo simulations with $M=500$ trials. One can see that the periodic behavior of both models is difficult to detect when analyzing only the time series. The periodicity is visible in the corresponding second-order statistic (i.e. variance).

\subsection{Compressor vibration  model}
\label{sec:compressor_model}

\begin{figure}
    \centering
    \includegraphics[width=0.6\linewidth]{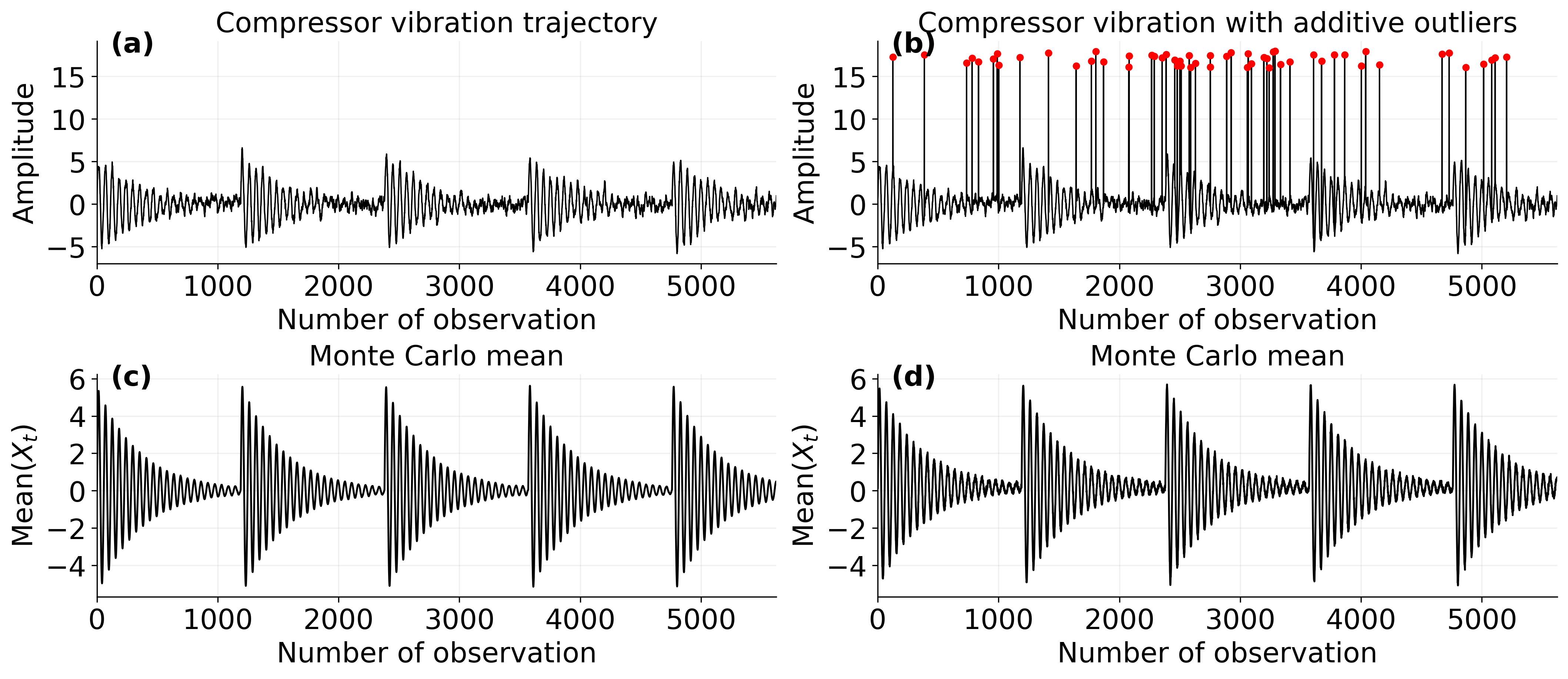}
    \caption{Example trajectories of (a): the compressor vibrations  and (b): its version with additive outliers described as $\tilde{B}_t= B_t$, together with the corresponding means estimated from M=500 Monte Carlo simulations (respectively: (c) and (d)).}
    \label{fig:traj_compressor}
\end{figure}

As a second cyclostationary model, we consider the time series given by the following equation \cite{grzesiek2026impulsivity}:
\begin{eqnarray}\label{model2}
    X_t=b(t)+\tilde{Z}_t,
\end{eqnarray}
where $b(t)$ is a deterministic function defined on the integers such that $b(t)=b(t+T)$ and $\{\tilde{Z}_t\}$ form a stationary time series. In general, the model defined in Eq. (\ref{model2}) fulfills conditions (\ref{eq:cyclostationary}); however, in this case, the mean function is periodic with a period of $T$, while the autocovariance function is constant with respect to $t$. Indeed, we have:
\begin{eqnarray}
    \mathbb{E}(X_t)=b(t)+\mathbb{E}\left(\tilde{Z}_t\right)=b(t)+c(0)=b(t+T)+c(0)= \mathbb{E}(X_{t+T}), \quad cov(X_t, X_{t+h})=cov(\tilde{Z}_t,\tilde{Z}_{t+h})=c(h),
\end{eqnarray}
where $c(h)$ is the autocovariance function of $\{\tilde{Z}_t\}$ and is independent of $t$.
When we analyze the model defined in Eq. (\ref{eq:general_signed}) with $\{X_t\}$ defined as in Eq. (\ref{model2}), the time series $\{Y_t\}$ possesses similar properties as discussed above. More precisely, we have:
\begin{eqnarray}
\mathbb{E}(Y_t)=b(t)+c(0)+\mathbb{E}(B)=\mathbb{E}(Y_{t+T}),    \quad cov(Y_t, Y_{t+h})=c(h)+\mathbb{I}_{h=0}\sigma^2_{B}.
\end{eqnarray}
Thus, $\{Y_t\}$ is also cyclostationary. In our analysis, to demonstrate the practical applicability of the proposed methodology, we consider a very specific $b(t)$ function in Eq. (\ref{model2}) that imitates the periodic components observable in a real vibration signal from a compressor, see e.g. \cite{grzesiek2026impulsivity}. Compressor signals represent a canonical example of cyclostationary industrial data, where periodic oscillations arise from the mechanical operation of rotating components. In our case, $b(t)$ represents periodic oscillatory responses generated by the compressor operation:
\begin{equation}
b(t) = \sum_{k=0}^{\infty} \tilde{b}(t - kT_p),
\label{eq:periodic_response}
\end{equation}
where $T_p$ is the period of the compressor cycle. The response generated by a single disturbance is given by:
\begin{equation}
\tilde{b}(t) = I \cdot e^{-t/(\tau_d f_s)} \sin\left(2\pi \frac{f_0}{f_s} t + \phi\right),
\label{eq:impulse_response}
\end{equation}
where $I$ is the excitation amplitude, $\tau_d$ is the decay time constant, $f_0$ is the oscillation frequency, $f_s$ is the sampling frequency, and $\phi$ is the initial phase. 

By such definition of the $b(t)$ function we have $T_p=1190$. In addition, we assume $\{\tilde{Z}_t\}$ is a stationary AR(2) model, i.e., a model defined as in Eq. (\ref{eq:PAR}) with $\kappa=2$ and $s_1(t), s_2(t)$ as appropriate constant functions (that assure the stationarity of the AR(2) model). In Fig. \ref{fig:traj_compressor} we demonstrate the example trajectory of the time series from model (\ref{model2}) with parameters: $f_s = 25000$ Hz, AR coefficients: $s_1(t)=s_1 = 0.6$, $s_2(t)=s_2 = -0.2$; $f_0 = 4000$ Hz, $\tau_d = 0.005$ s (left panel) and the corresponding model (\ref{eq:general_signed}) with $a=16$, $b=18$ and $p=0.01$ (right panel). In such a case we consider $\tilde{B}_t=B_t$. In the bottom panel we present the mean functions of both models received using the Monte Carlo simulation of $M=500$ trials. One can see that for this model, for both versions (with and without additive outliers) the periodic behavior is clearly visible both in time series representations and the mean functions. In the applied contamination parameterization, the number of non-cyclic impulses is relatively small, resulting in only a minor difference in the mean value. This difference is observable primarily during the stage of cyclic oscillation decay. If a larger number of impulses and/or impulses with higher amplitudes were considered, their effect on the change in the mean value would become more pronounced. This is because a greater number of impulses would increase the probability of their occurrence at similar spatial locations, thereby contributing more significantly to the variation of the mean value.

\section{Methodology for anomaly detection in cyclostationary signals}
\label{sec:methodology}

This section provides a brief overview of the baseline methodology, denoted as CALM (Continuous Anomaly Localization for univariate and Multivariate data), which enables the real-time identification of anomalous observations in multivariate data~\cite{witulska2026realtime}. Subsequently, we present an enhanced version of CALM, referred to as PeriodicCALM, which incorporates a cyclic envelope to account for dependencies associated with cyclostationarity of given signal. This extension enables the method to distinguish genuine anomalies from observations that deviate from the expected value range but remain consistent with the underlying cyclic pattern, thereby reducing false positive detections. Finally, we describe the proposed adaptation mechanism for anomaly detection models under changing environmental conditions.

\subsection{Anomaly detection based on kernel density estimation}
\label{sec:cp-method}

The baseline CALM method for anomaly detection - estimates the distribution of historical, non-anomalous observations using kernel density estimation (KDE) \cite{silverman2018density} and identifies observations associated with low estimated probability density. The resulting density-based scores are compared with data-driven thresholds obtained using a bootstrap procedure. A detailed description of the theoretical foundations and implementation of CALM is provided in~\cite{witulska2026realtime}.

\subsection{Enhanced methodology for cyclostationary signals}
\label{sec:periodic_calm}

\begin{figure}
    \centering
    \includegraphics[width=\linewidth]{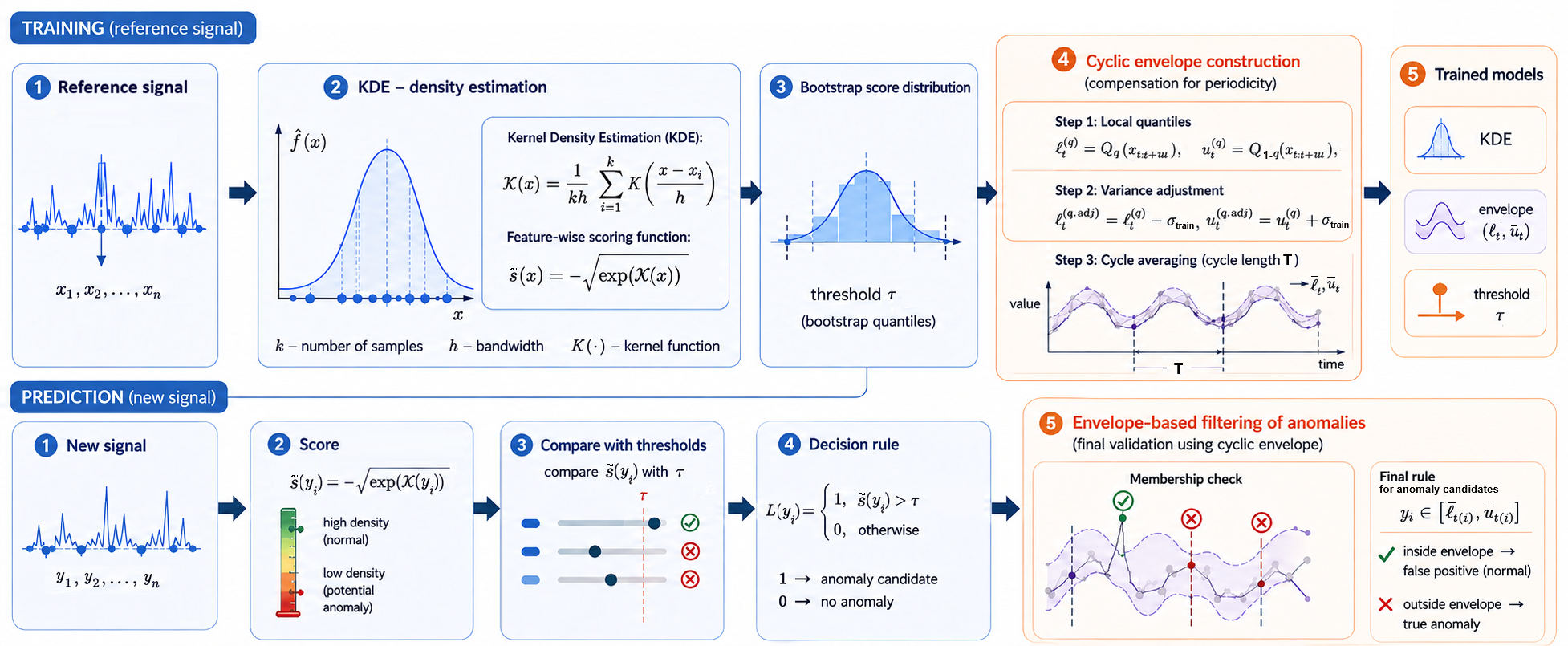}
    \caption{Schematic block of PeriodicCALM.}
    \label{fig:schematic_block}
\end{figure}

The baseline CALM method assumes independence between observations. However, in real systems signals are often periodic -- natural oscillations may be misclassified as anomalies, leading to significantly increased false positive rates. Examples of such systems include rotating machinery, compressors, and vibration-based sensors. This motivates the development of a post-processing mechanism to account for cyclic structure.

The enhanced method, denoted as PeriodicCALM, extends CALM by incorporating a cyclic envelope that models natural variability within each signal cycle. The key insight is that in cyclostationary signals, the acceptable range of values varies systematically with the phase position within each cycle. By constructing an envelope that captures this cycle-dependent variability, we can filter out false positives caused by normal cyclic behavior while retaining genuine anomalies. The method operates in two stages: (1) apply the baseline CALM algorithm to detect candidate anomalies, and (2) validate each detected point against the cyclic envelope to determine whether it represents normal cyclic behavior or a true anomaly.

PeriodicCALM employs the following input parameters.
The first group is inherited from the baseline CALM algorithm~\cite{witulska2026realtime}:
(i)~the number of bootstrap resamples $N_r$ used for dynamic threshold
estimation,
(ii)~the bootstrap quantile level $q_b$ applied during threshold estimation,
and
(iii)~the sample fraction $\eta$ serving as an auxiliary parameter in the
bootstrap process.
The second group consists of parameters specific to the cyclic-envelope
extension:
(iv)~the sliding-window width $w$ used for local quantile estimation
(Step~1 below), and
(v)~the quantile level $q$ defining the lower ($Q_q$) and upper ($Q_{1-q}$)
bounds of the local envelope (Step~1 below), and
(vi)~the known cycle length $T$ required for cycle averaging (Step~3 below),
equal to the cyclostationary period introduced in Section~\ref{sec:models}
(for the compressor model, $T=T_p$).

Let $x_1, \ldots, x_N \in \mathbb{R}$ denote the training signal without anomalies (a realization of $\{X_t\}$) and $y_1, \ldots, y_{N_{\mathrm{test}}} \in \mathbb{R}$ denote the new observations to be scored (a realization of $\{Y_t\}$). A step-by-step description of PeriodicCALM is presented below.

\begin{enumerate}
  \item \textbf{Training phase}
  \begin{enumerate}[(a)]
    \item Load the training data without anomalies $x_1, \ldots, x_N \in \mathbb{R}$.
    \item Calculate threshold $\tau$ as described in~\cite{witulska2026realtime}
          (CALM applied to the univariate signal, i.e.\ CALM with $d = 1$).
    \item Calculate the cyclic envelope as follows.
    \begin{itemize}
      \item \textbf{Step~1: Local quantile estimation.}
            For each time point $i$ in the training signal, compute local
            quantiles using a sliding window of width $w$:
            \begin{equation}
              \ell_{i}^{(q)} = Q_{q}(x_{i:i+w}),
              \qquad
              u_{i}^{(q)} = Q_{1-q}(x_{i:i+w}),
              \label{eq:local_quantiles}
            \end{equation}
            where $Q_{q}(\cdot)$ denotes the $q$-quantile and
            $x_{i:i+w}$ represents the signal values within the window
            starting at position~$i$, i.e., $x_{i:i+w} = x_i, x_{i+1},...,x_{i+w}$.

      \item \textbf{Step~2: Variance adjustment.}
            To account for estimation uncertainty and signal variability, the
            local bounds are adjusted by the training-data standard deviation
            $\sigma_{\mathrm{train}}$:
            
            \begin{equation}
              \ell_{i}^{(q,\mathrm{adj})} = \ell_{i}^{(q)} - \sigma_{\mathrm{train}},
              \qquad
              u_{i}^{(q,\mathrm{adj})} = u_{i}^{(q)} + \sigma_{\mathrm{train}},
              \label{eq:variance_adjustment}
            \end{equation}

            where
            \begin{equation}
             \sigma_{\mathrm{train}} = \sqrt{\tfrac{1}{N-1}\sum_{k=1}^{N}(x_k - \bar{x})^2}, \quad \bar{x} = \tfrac{1}{N}\sum_{k=1}^{N} x_k.
            \end{equation}

      \item \textbf{Step~3: Cycle averaging.}
            Given a known cycle length $T$, partition the signal into complete
            cycles and average the envelope bounds across all $C$ cycles:
            \begin{equation}
              \bar{\ell}_{t} = \frac{1}{C}\sum_{c=1}^{C}
                \ell_{t + (c-1)T}^{(q,\mathrm{adj})},
              \qquad
              \bar{u}_{t} = \frac{1}{C}\sum_{c=1}^{C}
                u_{t + (c-1)T}^{(q,\mathrm{adj})},
              \label{eq:cycle_averaging}
            \end{equation}
            where $C$ is the number of complete cycles in the training data
            and $t \in \{1, \ldots, T\}$ is the phase position within the
            cycle.
    \end{itemize}
  \end{enumerate}

  \item \textbf{Prediction phase}
  \begin{enumerate}[(a)]
    \item Load new data $y_1, \ldots, y_{N_{\mathrm{test}}} \in \mathbb{R}$.
          For each $m \in \{1, \ldots, N_{\mathrm{test}}\}$:
    \begin{itemize}
      \item Compute the anomaly score $\tilde{s}(y_m)$ as defined
            in~\cite{witulska2026realtime}.
      \item Assign the label:
            \begin{equation}
              L(y_m) =
              \begin{cases}
                1, & \text{if } \tilde{s}(y_m) > \tau
                     \ \text{ and } \
                     y_m \notin \bigl[\bar{\ell}_{t(m)},\,
                                      \bar{u}_{t(m)}\bigr], \\
                0, & \text{otherwise.}
              \end{cases}
            \end{equation}
    \end{itemize}
  \end{enumerate}
\end{enumerate}

As in the baseline CALM case, if the detected point lies inside the envelope ($y_m \in [\bar{\ell}_{t(m)}, \bar{u}_{t(m)}]$), it is reclassified as a false positive representing normal cyclic behaviour; only points falling outside the envelope are retained as true anomalies.

Fig.~\ref{fig:schematic_block} illustrates the schematic of PeriodicCALM for both the training and prediction phases. The steps highlighted in blue correspond to the baseline CALM procedure, whereas the steps highlighted in orange represent the extensions introduced by PeriodicCALM, which involve the construction and utilization of the periodic envelope.

The generalisation to multivariate signals ($d > 1$) can be achieved analogously to the basic CALM method by applying the procedure independently to each signal component and subsequently combining the resulting per-feature labels. However, here we present the complete procedure for the univariate case, as considering a single component allows us to evaluate the effectiveness of the cyclic-envelope filtering step in isolation, without introducing the additional complexity associated with combining feature-wise labels.

The novelty of the PeriodicCALM, in comparison to the baseline method, is related to two main points:
\begin{itemize}
  \item Cyclic envelope construction mechanism was added that models the phase-dependent variability of the signal using local quantiles, variance adjustment, and cycle averaging. This enables the method to distinguish between normal cyclic oscillations and genuine anomalies without requiring explicit seasonal decomposition or preprocessing.

  \item As a result, the enhanced PeriodicCALM method significantly reduces false alarms in periodic data while preserving the computational simplicity and real-time capability of the original CALM framework. The method improves robustness in industrial applications involving cyclostationary signals such as rotating machinery, compressors, and vibration-based monitoring systems.
\end{itemize}

\subsection{Adaptation of anomaly detection models to changing environmental conditions}

In complex systems, real-time anomaly detection model calibrated on representative anomaly-free data can reliably identify deviations as long as the underlying operating conditions remain stable. However, when a regime change (such as a shift in the signal distribution, a change in scale, or a modification of process parameters) occurs then the statistical characteristics of the observations may change and invalidate the previously fitted model. Without model adaptation, such changes can result in persistent false alarms, as the detector may incorrectly interpret the new normal operating regime as anomalous behavior. Therefore, anomaly detection should be complemented by a dedicated change-detection mechanism that continuously monitors the data for significant changes in the underlying process. Upon detecting a regime change, the system should temporarily suspend anomaly decisions and trigger model retraining using data representative of the new normal operating conditions before resuming anomaly detection. Additionally, if process-parameter changes are directly monitored, their occurrence can be used to schedule model retraining proactively.

\section{Simulation study}
\label{sec:simulation}

In this section, we evaluate the performance of the proposed anomaly detection method for two types of periodic signals (\textit{Model 1} and \textit{Model 2} described below).   
The two models allow the method to be assessed both under controlled simulation conditions and in a setting representative of industrial vibration data. All experiments were performed on a workstation featuring an \textit{Intel Xeon E5-2680 v4} processor and \textit{247 GB of RAM}, running \textit{Ubuntu 22.04.4 LTS} with Linux kernel version \textit{6.12.31-talos}. The computations were carried out exclusively on the CPU, without GPU acceleration. The reported results are therefore based on repeated independent realizations of the corresponding data-generating process.

\subsubsection*{\textit{Model 1}: PAR model with additive outliers}
The first model (\textit{Model 1}) is described by Eq. (\ref{eq:general_signed}) with $\{X_t$\} given as PAR(1) time series (described in Section~\ref{sec:model}) with anomalies described as $\tilde{B}_t=\operatorname{sign}(X_t) \cdot B_t$. The periodic function is given as $s_1(t)=s(t) = 0.5 + 0.3 \sin(2\pi t/T),$ and period $T=12$. The standard deviation of the innovation series is set to $\sigma=0.25$, and the total sample size is $N=4500$. 

\subsubsection*{\textit{Model 2}: Compressor vibration model with additive outliers}
The second model (\textit{Model 2}) considers a signal representative of industrial compressor vibration measurements. It is described by Eq. (\ref{eq:general_signed}) with $\{X_t\}$ given as a compressor vibration signal exhibiting pronounced periodic oscillations (described in Section~\ref{sec:compressor_model}) with anomalies described as $\tilde{B}_t= B_t$. 
The signal is sampled at $f_s=25000$ Hz and has a duration of 0.25 second (so $N=6250)$. Its stochastic component is described by an AR(2) model (i.e. PAR(2) model with $T=1)$ with coefficients $s_1(t)=s_1=0.6$ and $s_2(t)=s_2=-0.2$, while the periodic component is characterized by an oscillation frequency of $f_0=4000$ Hz and a decay constant of $\tau_d=0.005$ s. 

\subsubsection*{Contamination level and frequency}
For both models, 100 independent Monte Carlo replications are performed for each considered parameter configuration. The contamination is introduced through additive impulses, with the impulse probability taking values $p \in \{0.05,0.1\}$ for simulations corresponding to \textit{Model 1}, and $p \in \{0.005,0.01\}$ for simulations corresponding to \textit{Model 2}.

To unify the presentation of the results for both models, we introduce a reference scale, denoted D, that indicates how far the considered observations lie from the typical upper edge of still-standard variation. Specifically, D is obtained by sliding a window of length 100 along the contaminated series (from 100 simulations), computing the 0.95-quantile in each window, and averaging these local quantiles. The resulting value summarizes the magnitude of observations that remain consistent with ordinary behavior, even though they already occupy the upper tail of the local distribution. Expressing the impulse amplitudes relative to D therefore makes the two models comparable: it shows how far the added disturbances stand above this common, still-standard upper bound.

The impulse amplitude is generated within one of the following ranges: $
(a,b) \in \{(1.5D,4D), (1.5D, 8D), (4D, 6D),$ $(4D, 10D), (6D, 8D), (6D, 12D)\}.$
Thus, the experimental design covers different levels of both anomaly frequency and amplitude.

\subsubsection*{Parametrization of PeriodicCALM}
For the CALM component, the quantile filtering parameters are set to $(q_{\mathrm{low}},q_{\mathrm{high}})=(0.1,0.9)$, while the bandwidth of the kernel density estimator is determined using Silverman's rule. The bootstrap-based threshold estimation is performed using $N_r=100$ resamples, a sample fraction of $\eta=0.75$, and threshold quantile $q_b=0.99$. For the periodic envelope, the window size is set to $w=100$, and the envelope is widened by $\pm \sigma_{\mathrm{train}}$ to account for the variability of the training observations. For \textit{Model 1}, the cycle length is $T=12$, and the generated signal is divided into a training segment containing 1000 observations and a test segment of length $N_{\mathrm{test}}=3500$. For \textit{Model 2}, the cycle length is $T=T_p=1190$, and the generated signal is divided into a training segment containing 2500 observations and a test segment of length $N_{\mathrm{test}}=10000$.

\subsection{Performance assessment of the enhanced anomaly detection method compared to baseline}
\label{sec:efficacy}

\begin{figure}
    \centering
    \includegraphics[width=\textwidth]{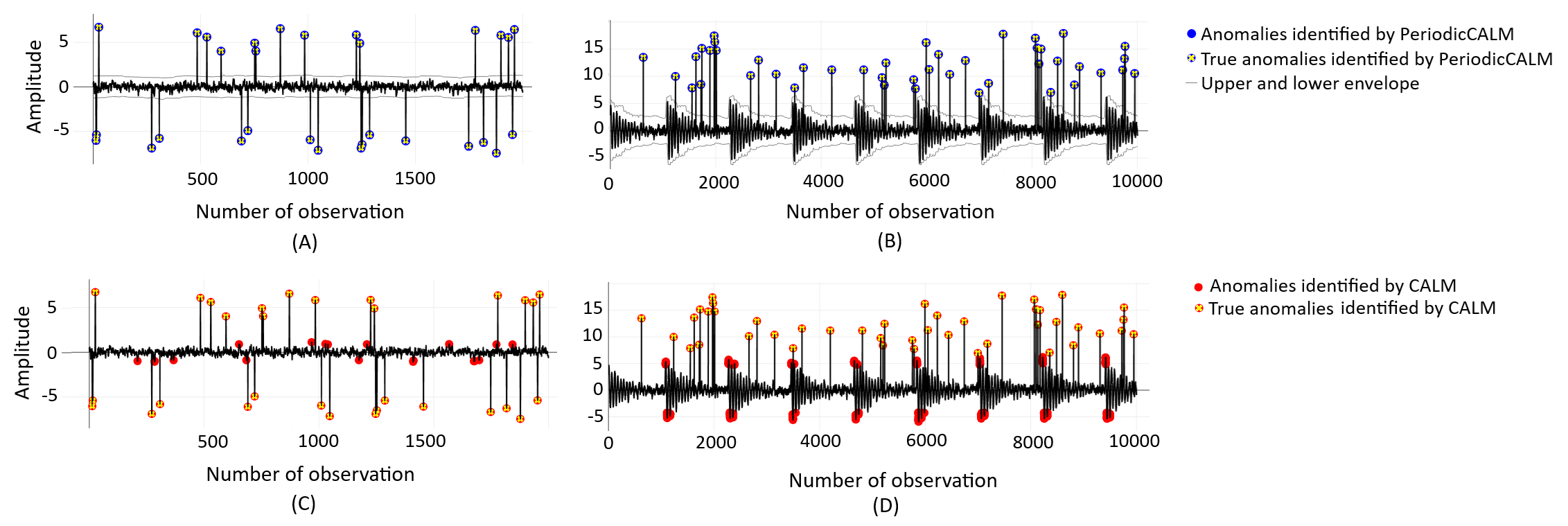}
    \caption{Illustrative examples for anomaly identification: (A) by PeriodicCALM method used for data from \textit{Model 1}, (B) by PeriodicCALM method used for data from \textit{Model 2}, (C) by the baseline CALM used for data from \textit{Model 1}, (D) by the baseline CALM used for data from \textit{Model 2}.}

    \label{fig:calm_comparison}
\end{figure}

In this section, we conduct a comparative analysis of the PeriodicCALM and CALM methods, with particular emphasis on the impact of the proposed periodic envelope extension on the reduction of false anomalies arising from the inherent periodicity of the signals, such as recurring impulses. A visualization of anomaly detection using the \textsc{CALM} and PeriodicCALM  methods for exemplary data generated according to the \textit{Model 1}, is presented in Fig.~\ref{fig:calm_comparison}(a) and (c). In contrast, Fig.~\ref{fig:calm_comparison}(b) and (d) presents the corresponding visualization for exemplary data generated according to the \textit{Model 2}. The ground truth (i.e., real anomalies) as marked by yellow "x". As can be seen, the baseline method detects numerous false positives corresponding to normal cyclic peaks, while PeriodicCALM effectively filters these out.

The subsequent analysis focused on a quantitative assessment of the performance of the considered methods using extensive Monte Carlo simulations over a broad parameter grid. The evaluation of the proposed anomaly identification method can be formulated as a binary classification problem with imbalanced classes \cite{hossain2024evaluating}. In this setting, anomalous observations constitute the minority class, whereas observations that do not deviate from the expected behavior form the majority class. This class imbalance is a characteristic feature of anomaly detection problems and should be explicitly considered when evaluating the performance of the proposed method. In particular, accuracy alone may provide a misleading assessment, as a classifier can achieve a high accuracy by predominantly assigning observations to the majority class.

In general, the evaluation of classification models is most commonly based on the analysis of a confusion matrix, which consists of four basic elements: true positives (TP), true negatives (TN), false positives (FP), and false negatives (FN) \cite{chicco2020advantages}. Based on these quantities, a number of performance measures can be defined, allowing for a comprehensive assessment of the classifier. Let $g \in \{0,1\}$ denote the true class label, where $1$ corresponds to an anomalous observation. The performance of the proposed PeriodicCALM method and CALM method were evaluated using standard classification metrics~\cite{chicco2020advantages}. 

Most measures dedicated to quantify the performance of classification, including accuracy, precision, recall, F1-score, and specificity, attain their optimal values close to $1$, whereas error measures, such as the false positive rate (FPR) and false negative rate (FNR), should be as close to $0$ as possible. The Matthews correlation coefficient (MCC) ranges from $-1$ to $1$, where values close to $1$ indicate very good classification performance, a value close to $0$ corresponds to performance comparable to random classification, and values close to $-1$ indicate a classification that is opposite to the true labels. Due to the imbalanced nature of the considered problem, particular attention is paid to measures such as precision, recall, F1-score, and MCC, which provide more informative assessments than accuracy alone.

\begin{figure}
    \centering

    \begin{subfigure}{0.99\textwidth}
        \centering
        \includegraphics[width=0.9\textwidth]{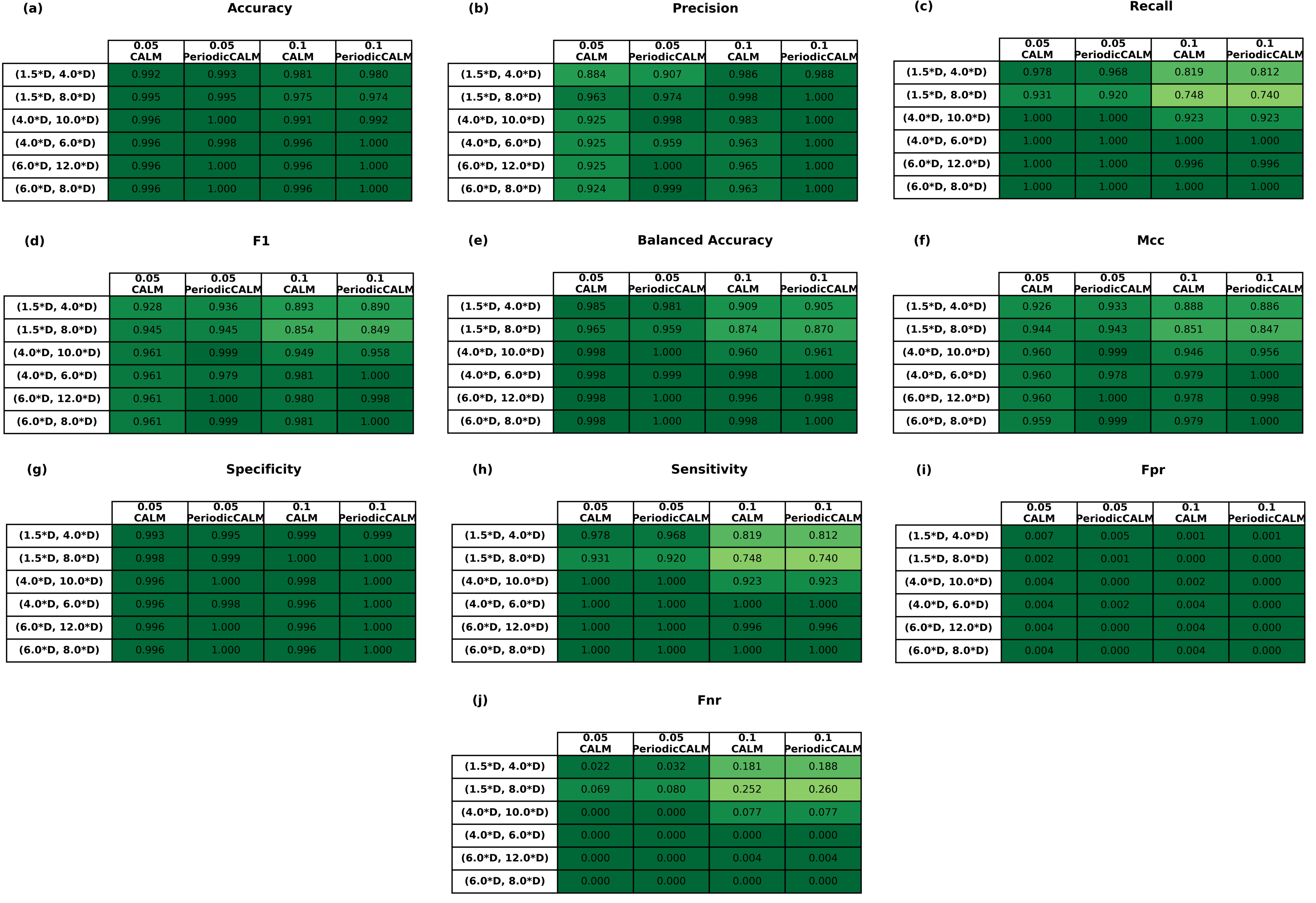}
        \caption{Results for data generated according to the \textit{Model 1}.}
        \label{fig:metrics_heatmaps_par}
    \end{subfigure}

    \vfill

    \begin{subfigure}{0.99\textwidth}
        \centering
        \includegraphics[width=0.9\textwidth]{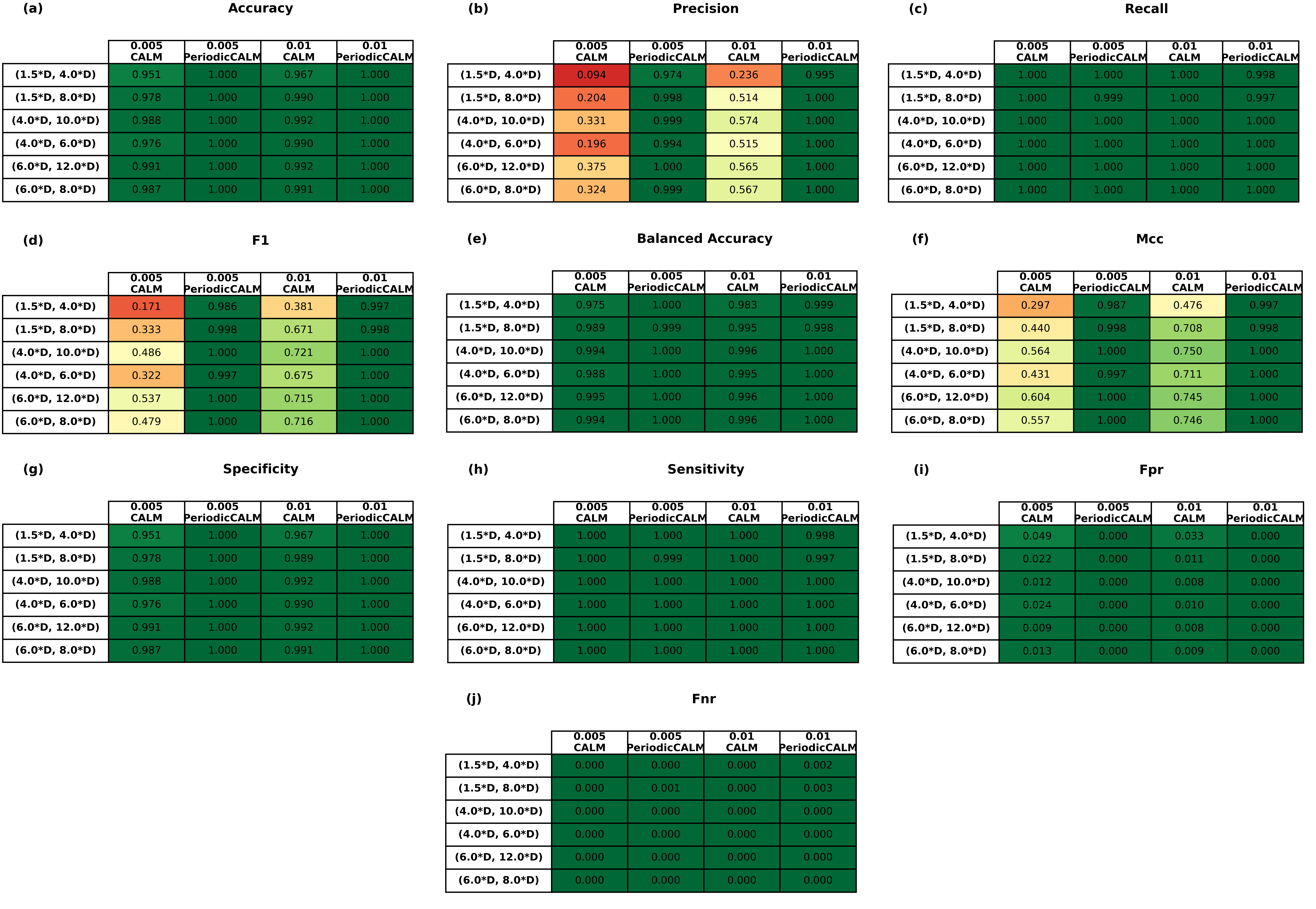}
        \caption{Results for data generated according to the \textit{Model 2}.}
        \label{fig:metrics_heatmaps_compressor}
    \end{subfigure}

    \caption{Comparative analysis of CALM and PeriodicCALM performance across different impulse configurations for simulated data. The first column defines the intervals for the $(a,b)$ values used in the simulations. Here, $D=4.5$. The values of $p$ and the name of the method used to compute the corresponding metric are indicated in the table headers.}
    \label{fig:metrics_heatmaps}
\end{figure}

The comparative analysis of the methods' accuracy is presented in Fig.~\ref{fig:metrics_heatmaps_par}, showing heatmaps for all evaluation metrics across different impulse parameter configurations for data generated according to the \textit{Model 1}, and corresponding results for data generated according to the \textit{Model 2}. Additionally, Tab.~\ref{tab:par_tab_combined} reports complementary comparison measures for corresponding models in the form of differences, defined as $\Delta = V_{\mathrm{PeriodicCALM}} - V_{\mathrm{CALM}}$, where $V$ denotes the corresponding performance quantity (FP, TP, FN, TN) obtained for PeriodicCALM and CALM, respectively, and percentage differences, defined as $\Delta / V_{\mathrm{CALM}} \times 100\%$.

The simulation results demonstrate that the proposed PeriodicCALM method improves anomaly detection performance for cyclic data compared with the baseline CALM approach. As one can see, in Tab.\ref{tab:par_tab_combined} - the incorporation of cyclic envelope filtering substantially reduces FP - especially for larger period T and stronger dependencies in data (like for data generated from \textit{Model 2}). At the same time, the additional filtering step preserves a high recall, as genuine non-cyclic impulses occurring outside the learned cyclic envelope are correctly identified. We can observe it by analyzing Tab.\ref{tab:par_tab_combined}, where mean $\Delta$ for TP is equal to 0 for almost all configuration of impulses.  

Interestingly, the analysis of the metrics presented in Fig.~\ref{fig:metrics_heatmaps_par} confirms that, for the data generated by \textit{Model 1}, the improvement in performance is not as pronounced. This suggests that, to some extent, CALM is capable of handling data characterized by weak dependencies. However, the metrics presented in Fig.~\ref{fig:metrics_heatmaps_compressor} clearly demonstrate that, in the presence of strong dependencies, incorporating the cyclic envelope substantially reduces the number of FPs, thereby improving both precision and F1-score. In the most extreme case, precision increases from 0.094 to 0.974, while the F1-score improves from 0.171 to 0.986.

The proposed method also exhibits robustness with respect to the characteristics of the contamination. Similar performance is observed for different impulse probability levels, $p$, as well as for the considered amplitude ranges. These results indicate that the method remains effective under varying degrees of anomaly occurrence and magnitude. 
\begin{table}[p]
\centering
\caption{Comparative analysis of CALM and PeriodicCALM performance across different impulse configurations for simulated data. The first two columns define the intervals for the $(a,b)$ values used in the simulations. Results for Model 1 and Model 2 are presented separately, with $D=0.5$ and $D=4.5$, respectively. The values of $p$ and the name of the method used to compute the corresponding metric are indicated in the table headers.}
\label{tab:par_tab_combined}
\resizebox{\textwidth}{!}{%
\begin{tabular}{cc|c|rrrrrrrr|c|rrrrrrrr}
\toprule
\multirow{3}{*}{$a/D$} &
\multirow{3}{*}{$b/D$}
&
\multicolumn{9}{c|}{\textbf{Model 1} ($D=0.5$)}
&
\multicolumn{9}{c}{\textbf{Model 2} ($D=4.5$)}
\\
\cmidrule(lr){3-11} \cmidrule(l){12-20}

&
&
\multirow{2}{*}{$p$}
&
\multicolumn{2}{c}{TP}
&
\multicolumn{2}{c}{TN}
&
\multicolumn{2}{c}{FN}
&
\multicolumn{2}{c|}{FP}
&
\multirow{2}{*}{$p$}
&
\multicolumn{2}{c}{TP}
&
\multicolumn{2}{c}{TN}
&
\multicolumn{2}{c}{FN}
&
\multicolumn{2}{c}{FP}
\\
\cmidrule(lr){4-5}
\cmidrule(lr){6-7}
\cmidrule(lr){8-9}
\cmidrule(lr){10-11}
\cmidrule(lr){13-14}
\cmidrule(lr){15-16}
\cmidrule(lr){17-18}
\cmidrule(l){19-20}

&
&
&
\% & $\Delta$
&
\% & $\Delta$
&
\% & $\Delta$
&
\% & $\Delta$
&
&
\% & $\Delta$
&
\% & $\Delta$
&
\% & $\Delta$
&
\% & $\Delta$
\\
\midrule

\multirow{2}{*}{1.5}
&
\multirow{2}{*}{4}
&
0.05
& $-0.78$ & $-1$
& $0.08$ & $2$
& $25.00$ & $1$
& $-14.29$ & $-2$
&
0.005
& $0.00$ & $0$
& $5.17$ & $488.5$
& -- & $0$
& $-100.00$ & $-488.5$
\\
&
&
0.10
& $0.00$ & $0$
& $0.00$ & $0$
& $0.00$ & $0$
& $0.00$ & $0$
&
0.010
& $0.00$ & $0$
& $3.39$ & $324.5$
& -- & $0$
& $-100.00$ & $-324.5$
\\
\addlinespace

\multirow{2}{*}{1.5}
&
\multirow{2}{*}{8}
&
0.05
& $0.00$ & $0$
& $0.00$ & $0$
& $0.00$ & $0$
& $-25.00$ & $0$
&
0.005
& $0.00$ & $0$
& $2.13$ & $207.5$
& -- & $0$
& $-100.00$ & $-207.5$
\\
&
&
0.10
& $0.00$ & $0$
& $0.00$ & $0$
& $0.00$ & $0$
& $-100.00$ & $0$
&
0.010
& $0.00$ & $0$
& $1.08$ & $105.5$
& -- & $0$
& $-100.00$ & $-105.5$
\\
\addlinespace

\multirow{2}{*}{4}
&
\multirow{2}{*}{6}
&
0.05
& $0.00$ & $0$
& $0.21$ & $5$
& -- & $0$
& $-45.45$ & $-5$
&
0.005
& $0.00$ & $0$
& $2.47$ & $239.5$
& -- & $0$
& $-100.00$ & $-239.5$
\\
&
&
0.10
& $0.00$ & $0$
& $0.41$ & $9$
& -- & $0$
& $-100.00$ & $-9$
&
0.010
& $0.00$ & $0$
& $1.00$ & $98$
& -- & $0$
& $-100.00$ & $-98$
\\
\addlinespace

\multirow{2}{*}{4}
&
\multirow{2}{*}{10}
&
0.05
& $0.00$ & $0$
& $0.42$ & $10$
& -- & $0$
& $-100.00$ & $-10$
&
0.005
& $0.00$ & $0$
& $1.10$ & $108$
& -- & $0$
& $-100.00$ & $-108$
\\
&
&
0.10
& $0.00$ & $0$
& $0.18$ & $4$
& $0.00$ & $0$
& $-100.00$ & $-4$
&
0.010
& $0.00$ & $0$
& $0.76$ & $74.5$
& -- & $0$
& $-100.00$ & $-74.5$
\\
\addlinespace

\multirow{2}{*}{6}
&
\multirow{2}{*}{8}
&
0.05
& $0.00$ & $0$
& $0.42$ & $10$
& -- & $0$
& $-100.00$ & $-10$
&
0.005
& $0.00$ & $0$
& $1.20$ & $117.5$
& -- & $0$
& $-100.00$ & $-117.5$
\\
&
&
0.10
& $0.00$ & $0$
& $0.44$ & $10$
& -- & $0$
& $-100.00$ & $-10$
&
0.010
& $0.00$ & $0$
& $0.83$ & $81.5$
& -- & $0$
& $-100.00$ & $-81.5$
\\
\addlinespace

\multirow{2}{*}{6}
&
\multirow{2}{*}{12}
&
0.05
& $0.00$ & $0$
& $0.42$ & $10$
& -- & $0$
& $-100.00$ & $-10$
&
0.005
& $0.00$ & $0$
& $0.87$ & $85.5$
& -- & $0$
& $-100.00$ & $-85.5$
\\
&
&
0.10
& $0.00$ & $0$
& $0.40$ & $9$
& $0.00$ & $0$
& $-100.00$ & $-9$
&
0.010
& $0.00$ & $0$
& $0.80$ & $79$
& -- & $0$
& $-100.00$ & $-79$
\\

\bottomrule
\end{tabular}%
}
\end{table}

The comparative analysis shows that, for cyclostationary signals, the baseline CALM method tends to classify regular cyclic peaks as anomalies, leading to an elevated false positive rate and reduced specificity. PeriodicCALM alleviates this limitation by incorporating cycle-dependent variability into the decision process, thereby distinguishing regular periodic fluctuations from genuine anomalous deviations more effectively. The improvement of precision being particularly pronounced for smaller impulse amplitudes (see, Fig.~\ref{fig:metrics_heatmaps}). 
\clearpage\newpage

\subsection{Runtime computational complexity and memory}
\label{sec:complexity}

This section reports an empirical benchmark of the proposed procedure in terms of runtime and peak memory usage (measured by Python modules: \textit{time} and \textit{tracemalloc}, respectively). Such aspects are especially important in real-world applications, where low latency and minimal resource consumption are critical for real-time deployment. The experiments were repeated over 25 Monte Carlo trials for each tested configuration, and the reported values correspond to averages across trials. In the terminology used here, the training phase refers to the fit-and-predict part based on the baseline CALM detector, while envelope construction refers to stage in  which comprise the cyclic envelope estimation and the subsequent rule-based filtering step. The goal of this benchmark is to quantify the computational cost of both components separately and to assess how it varies with the training set size and the total sample length.

The mean runtime and its corresponding standard deviation, and the mean peak memory usage and its standard deviation, is presented in Tab.~\ref{tab:runtime_memory_benchmark}. The results indicate that the training stage is the dominant computational component in terms of both runtime and memory usage across the tested configurations. Importantly, its cost is not driven by the total sample length $N$, but rather by the amount of data required to represent several full cycles, with we recommend to use at least three complete cycles for training. The envelope construction and rule-based filtering contribute only marginally to the overall execution time, remaining below one second in all cases. A similar pattern is observed for memory consumption, where the training stage accounts for the majority of the peak footprint. Therefore, the practical cost is determined primarily by the need to provide a sufficiently long training segment to capture a few complete cycles, while the overall length of the sample has little effect as long as this requirement is satisfied. For very long cycles, preprocessing that reduces the data length may still be beneficial, whereas for typical cases the base CALM prediction and cyclic-envelope filtering remain fast and efficient.

\begin{table}
\caption{Mean $\pm$ standard deviation of runtime and peak memory over 25 Monte Carlo trials. Runtime is reported in seconds and peak memory in MB.}
\centering
\resizebox{\textwidth}{!}{%
\begin{tabular}{rr|ccc|ccc}
\toprule
&
&
\multicolumn{3}{c|}{\textbf{Runtime [s]}}
&
\multicolumn{3}{c}{\textbf{Peak memory [MB]}}
\\
\cmidrule(lr){3-5}
\cmidrule(l){6-8}

\textbf{Training size}
&
\textbf{Stage}
&
$N=1000$ & $N=2000$ & $N=3000$
&
$N=1000$ & $N=2000$ & $N=3000$
\\
\midrule

100
& training phase
& 7.181 $\pm$ 0.619
& 7.676 $\pm$ 0.604
& 6.562 $\pm$ 1.981
& 567.249 $\pm$ 10.152
& 567.029 $\pm$ 10.047
& 565.395 $\pm$ 8.743
\\

100
& envelope construction
& 0.008 $\pm$ 0.002
& 0.021 $\pm$ 0.023
& 0.042 $\pm$ 0.032
& 172.426 $\pm$ 6.569
& 333.398 $\pm$ 6.753
& 492.483 $\pm$ 5.890
\\
\addlinespace

500
& training phase
& 32.886 $\pm$ 2.836
& 31.524 $\pm$ 2.819
& 32.212 $\pm$ 2.631
& 2855.762 $\pm$ 30.076
& 2858.898 $\pm$ 44.469
& 2841.186 $\pm$ 29.025
\\

500
& envelope construction
& 0.496 $\pm$ 0.178
& 0.526 $\pm$ 0.123
& 0.518 $\pm$ 0.123
& 115.720 $\pm$ 5.068
& 277.145 $\pm$ 4.784
& 439.671 $\pm$ 4.617
\\
\addlinespace

1000
& training phase
& --
& 48.570 $\pm$ 20.864
& 40.574 $\pm$ 14.775
& --
& 5663.832 $\pm$ 42.704
& 5679.086 $\pm$ 25.993
\\

1000
& envelope construction
& --
& 0.753 $\pm$ 0.326
& 0.541 $\pm$ 0.013
& --
& 212.894 $\pm$ 5.008
& 370.355 $\pm$ 1.718
\\
\addlinespace

2000
& training phase
& --
& --
& 69.802 $\pm$ 16.404
& --
& --
& 11339.437 $\pm$ 36.526
\\

2000
& envelope construction
& --
& --
& 1.113 $\pm$ 0.022
& --
& --
& 240.599 $\pm$ 1.313
\\

\bottomrule
\end{tabular}%
}
\label{tab:runtime_memory_benchmark}
\end{table}

\section{Application to real data}
\label{sec:real_data}

To demonstrate the practical applicability of the proposed method, we consider real vibration data from compressor monitoring. Compressor signals represent a canonical example of cyclostationary industrial data where periodic oscillations arise from the mechanical operation of rotating components.

The measurements were collected by a condition monitoring system installed on a reciprocating compressor operating in the oil and gas industry. The primary function of the compressor is to increase the pressure of natural gas before its transportation through a pipeline to a gas-fired power generation facility. The machine is driven by a 2-MW electric motor and incorporates four compression cylinders, which raise the gas pressure progressively through multiple compression stages. The vibration measurements were acquired at a sampling frequency of 25 kHz using a sensor mounted in the vertical direction on one of the compressor cylinders.
The same datasets were analyzed in \cite{grzesiek2026impulsivity}, where the impulsive behavior of the signals was discussed.

The presented real-world example corresponds to the cyclostationary model with additive outliers described in Eq. (\ref{eq:general_signed}), where $\{X_t\}$ is modeled as described in Section~\ref{sec:compressor_model}. It is worth noting that, in other applications, anomalies may also occur with negative signs; however in this case we assume they take only positive values (see the discussion in Section \ref{sec:models} for more details). 

\begin{figure}    \centering
    \includegraphics[width=0.85\textwidth]{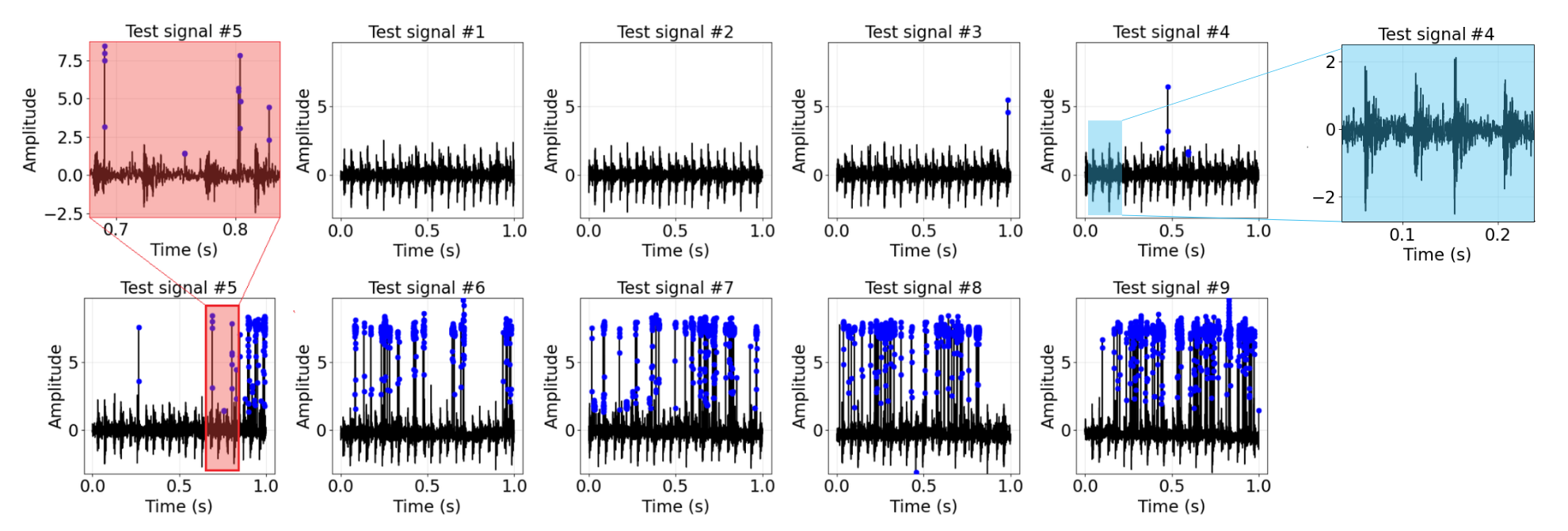}
    \caption{Example compressor signal with anomalies detected by PeriodicCALM. Only genuine non-cyclic disturbances are flagged as anomalies, while normal periodic oscillations are correctly classified as normal behavior.}
    \label{fig:periodic_calm_results}
\end{figure}

In this study, nine 1-s vibration records acquired by the same sensor at different time intervals are analyzed. The corresponding signals are presented in Fig.~\ref{fig:periodic_calm_results}. Signals \#1 and \#2 contain only periodic impulsive components associated with the normal operating conditions of the compressor. No additional non-periodic impulses are observed in these records (in the considered analysis, such impulses are treated as anomalies). The magnified signal segment indicated by the blue box in the upper-right corner of Fig.~\ref{fig:periodic_calm_results} provides a more detailed view of the waveform and demonstrates that the periodic impulses exhibit the characteristics of decayed harmonic oscillations. This observation provides a rationale for applying the proposed methodology to anomaly detection in dependent data, where the underlying signal structure is governed by recurrent oscillatory behavior.

Signals \#3 and \#4 contain individual non-periodic impulses, which are classified as anomalous events. In contrast, Signals \#5--\#9 exhibit a progressively increasing number of non-periodic impulses. The number of anomalous events therefore increases systematically with the signal index, providing a set of records with different levels of anomaly occurrence and enabling the performance of the proposed method to be evaluated under increasingly challenging conditions. Here, anomalies (non-periodic impulses observed in signals) can be attributed to e.g., electrical interference. Such disturbances require detection as they can significantly impact Key Performance Indicator (KPI) values, thereby distorting their interpretation. This may lead to erroneous decision-making during the overall system monitoring process. Upon detecting these anomalies, two primary approaches can be considered: either removing them from the data to prevent KPI distortion or replacing them with appropriate values. Alternatively, information regarding these anomalies can be reported during system monitoring to avoid false interpretations, such as indicating a non-existent system failure.

\begin{figure}
    \centering
    \includegraphics[width=0.7\textwidth]{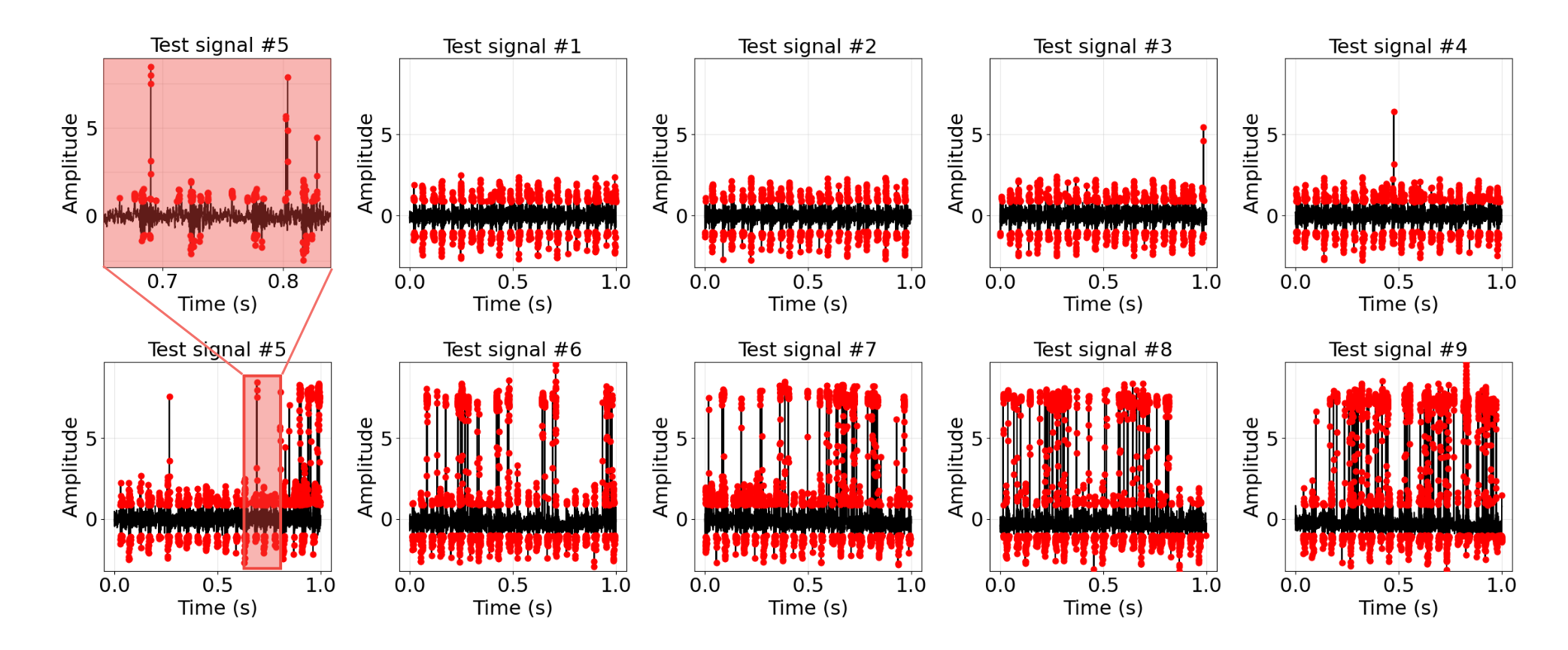}
    \caption{Example compressor signal with anomalies detected by baseline CALM.}
    \label{fig:calm_results}
\end{figure}

The results of anomaly detection using PeriodicCALM are shown in Fig.~\ref{fig:periodic_calm_results}. This  figure demonstrates the method's ability to identify genuine anomalies while ignoring normal cyclic behavior. The analogous result for the anomaly detection by baseline CALM method is presented in Fig.~\ref{fig:calm_results}. As shown, CALM tends to interpret cyclic impulses as anomalies. This behavior is implementation‑wise reasonable, since these observations indeed “fall outside the bounds” relative to the majority of signal samples. However, from an interpretive point of view they are not critical, as they signify normal device operation. Fig.~\ref{fig:periodic_calm_results} demonstrates that PeriodicCALM substantially reduces the number of detected anomalies. In fact, PeriodicCALM accepts approximately 35\% of the anomalies identified by CALM across the entire signal set. As illustrated, for Test signal \#1 and \#2, PeriodicCALM behaves as expected, reducing the anomaly count to zero, and for signals Test signal \#3 and \#4, it indicates only single anomalies in the form of non-cyclical pulses. For the remaining signals, as expected, it indicates a larger number of anomalies, reducing the occurrence of cyclical oscillations in the final results.

Hence, visual inspection confirms that the anomalies detected by PeriodicCALM are well justified aganist the anomalies detected by baseline CALM. However, one could consider whether the use of an additional threshold-based method would also reduce false positives. More importantly, example presented in Fig.~\ref{fig:periodic_calm_results} highlights the advantage of PeriodicCALM over conventional threshold-based methods. For instance, consider setting a threshold of 2.5 and applying a simple rule that labels an observation as anomalous whenever its value exceeds this threshold. The threshold-based approach has two major limitations. First, the threshold must be manually adjusted to the characteristics of the signal, which reduces its generality and requires additional domain knowledge. Second, a global threshold may fail to detect anomalies occurring in regimes with lower signal ranges. For example, the anomaly highlighted in the blue box in the upper-left corner in signal \#4, at approximately 0.75 s, would remain undetected because its magnitude does not exceed the global threshold. In contrast, PeriodicCALM can identify such regime-dependent anomalies by accounting for the local characteristics of the signal. Thus, PeriodicCALM allows for a trade-off between reducing false positives (related to cyclical patterns) while not over-rejecting true anomalies (as is the case with threshold-based methods).

The results confirm that PeriodicCALM is particularly suitable for continuous monitoring systems involving cyclic data, such as clinical decision support systems and predictive maintenance in industrial machinery.

\section{Summary and conclusions}
\label{sec:conclusions}

This work addressed the unresolved problem of real-time anomaly detection in cyclostationary signals, with particular emphasis on irregular, impulsive, and non-Gaussian anomalies. Such signals are characterized by strong phase-dependent dependencies and periodic variations that may be incorrectly identified as anomalies by conventional detection methods. To address this challenge, PeriodicCALM was proposed as an extension of the baseline CALM algorithm, redefined for dependent variables and augmented with a cyclic envelope construction and filtering mechanism. The proposed approach exploits the repetitive structure of cyclostationary signals to distinguish normal cyclic variations from genuine non-cyclic anomalies.

The experimental results demonstrate that PeriodicCALM substantially improves anomaly detection performance for cyclic data while maintaining a low false positive rate. In particular, cyclic envelope filtering effectively suppresses false alarms caused by normal periodic oscillations without compromising the detection of genuine non-cyclic anomalies. The most pronounced improvement increases precision from 0.094 to 0.974 and the F1-score from 0.171 to 0.986. These results demonstrate that the proposed filtering mechanism can markedly improve detection reliability, particularly for signals exhibiting strong dependencies and low-amplitude non-cyclic impulses. Its practical applicability is further supported by experiments on compressor vibration data, where PeriodicCALM provides an effective solution for industrial condition monitoring. Notably, approximately 65\% of the anomalies detected by the basic CALM method are rejected by PeriodicCALM as false indications associated with the cyclic operation of the machine.

The proposed methodology is also computationally efficient and introduces only minimal computational overhead relative to the baseline CALM algorithm. Runtime and memory measurements indicate that the additional operations required for cyclic envelope construction and filtering result in only a slight increase in computational cost, with the difference being practically negligible. This computational efficiency makes PeriodicCALM suitable for real-time monitoring applications.

Despite these advantages, several directions remain for further development. In the current implementation, the cycle length $T$ is specified by the user. Future work should therefore investigate its automatic estimation from data using spectral analysis, autocorrelation, or related signal-processing techniques, see, e.g., \cite{ZULAWINSKI2023115131}. Further research is also required to improve robustness to data drift, sensor outages, missing transmissions, and evolving data distributions. The incorporation of event-driven retraining mechanisms represents another promising direction, enabling the detection framework to adapt automatically to significant changes in system behavior.

Moreover, extending PeriodicCALM to signals exhibiting multiple or non-integer periodicities could further increase its applicability to complex industrial systems. Although the proposed methodology was validated in the context of cyclostationary signal processing, its underlying principles are not domain-specific and may therefore be applicable to a broader range of real-time anomaly detection problems involving dependent and temporally structured data.

\section*{Acknowledgments}

This work is supported by the NCN Weave-Unisono project entitled ``Advanced signal processing techniques for cyclostationary modelling in Gaussian and non-Gaussian noisy environment --- detection of cyclic sources, estimation, optimisation of algorithms and validation in the context of fault identification'' (No.~2025/07/Y/ST8/00070).

\section*{Conflict of interest}
No conflicts of interests.

\bibliographystyle{elsarticle-num}
\bibliography{mybibliography}

\end{document}